# Stage Division of Urban Growth Based on Logistic Model of Fractal Dimension Curves

Yanguang Chen

(Department of Geography, College of Urban and Environmental Sciences, Peking University, 100871, Beijing, China. Email: chenyg@pku.edu.cn)

**Abstract:** The time series of fractal dimension values of urban form always take on sigmoid curves. The basic model of these curves is logistic function. From the logistic model of fractal dimension curves, we can derive the growth rate formula and acceleration formula of city development. Using the inflexions of the fractal parameter curves, we can identify the different phases of urban evolution. The main results are as follows. (1) Based on the curve of fractal dimension of urban form, urban growth can be divided into four stages: initial slow growth, accelerated fast growth, decelerated fast growth, and terminal slow growth. The three dividing points are $0.2113D_{max}$, $0.5D_{max}$, and $0.7887D_{max}$, where $D_{max}$ is the capacity of fractal dimension. When the fractal dimension reaches half of its capacity value, $0.5D_{max}$, the urban growth rate reaches its peak. (2) Based on the curve of fractal dimension odds, urban growth can also be divided into four stages: initial slow filling, accelerated fast filling, decelerated fast filling, terminal slow filling. The three dividing points are $0.2113Z_{max}$, $0.5Z_{max}$, and $0.7887Z_{max}$, where $Z_{max}=/(2-D_{max})$ denotes the capacity of fractal dimension odds ($D_{max}<2$). Empirical analyses show that the first scheme based fractal dimension is suitable for the young cities and the second scheme based on fractal dimension odds can be applied to mature cities. A conclusion can be reached that logistic function is one of significant model for the stage division of urban growth based on fractal parameters of cities. The results of this study provide a new way of understanding the features and mechanism of urban phase transition.

# 1 Introduction

Urban growth is not a process of uniform change, but a dynamic process of slow change first, then fast change, and finally slow change. This means that the development of the city has stage property. This complex process involves urban phase transition and spatial replacement dynamics. Using different measurements to describe a city, we can find different characteristics of stages. Population is a basic measure of city size, which can reflect the growing stages of cities. Unfortunately, the census data with high reliability is not easy to obtain. An alternative approach is to measure the size of urban land use by means of remote sensing images. However, the measurement result of urban land use depend on the linear size of measurement scale. The scale-dependence of urban land area influence the effect of spatial analysis of urban growth and form. Fractal geometry is one of powerful tools for scale-free analysis of geographical phenomena. Fractal dimension is an index of spatial distribution and space filling of cities (Batty and Longley, 1994; Frankhauser, 1994). Using fractal dimension values, we can characterize the form of a city, and using the time series of fractal dimension values, we can describe urban growth from the scaling perspective (Benguigui *et al*, 2000; Frankhauser, 1994; Man and Chen, 2020; Manrubia *et al*, 1999; Murcio *et al*, 2015; Shen, 2002; Sun and Southworth, 2013).

The time series of fractal dimension values of urban form indicates new ways of identifying the development phases of a city. The key is to derive the general models for describing urban growth and determining the breakpoints of stage division. Generally speaking, fractal dimension values come between the topological dimension of a fractal and the Euclidean dimension of the embedding space in which the fractal is defined. This suggests that there is a squashing effect in the course of fractal dimension change, and the fractal dimension curves of urban form can be modeled with sigmoid functions (Chen, 2012; Chen, 2018). The basic expression of sigmoid functions is the well-known logistic function. Using the knowledge of calculus, we can deduce the models of urban growth rate and acceleration from the logistic function. These formulae can be applied to the stage division of urban growth. There are two research objectives in this paper: one is to develop the mathematical models of urban growth stage in theory; the other is to illustrate the method of stage division of urban transition based on fractal dimension in experience. These two goals complement each other. The rest of this paper is arranged as follows. In Section 2, the formulae of the dividing points of urban growth are derived from the logistic model of fractal dimension increase curve of urban form. In Section 3, four cities, London, Baltimore, Tel Aviv, and Shenzhen are employed to make case studies of urban stage division. In Section 4, several related questions are discussed, and finally, in Section 5, the study is concluded with a summary of main points.

## 2 Models

### 2.1 Fractal dimension curves of urban growth

In terms of scaling viewpoint, urban growth can be reflected by fractal dimension increase of urban form. Fractal dimension growth can be measured by time series of fractal dimension values. A time series of fractal dimension of urban form comprises unlimited continuous observed values of fractal parameters. Due to squashing effect of fractal dimension increase, a time series of fractal dimension values takes on a sigmoid curve (Chen, 2018). By analogy with the concept of *urbanization curve* (Davis, 1965; Cadwallader, 1996; Northam, 1979; Pacione, 2009), the sigmoid curves of fractal dimension series can be termed *fractal dimension curve* (Chen, 2018). The family of sigmoid functions include many different mathematical expressions, among which the basic and most important one is the logistic function. Logistic function is often treated as the representative of sigmoid functions (Mitchell, 1997). The logistic model of fractal dimension curves of urban growth can be expressed as

$$D(t) = \frac{D_{max}}{1+(D_{max}/D_0-1)e^{-kt}}, \qquad (1)$$

where $D(t)$ denotes the fractal dimension of urban form at time $t$, $D_0$ refers to the initial value of fractal dimension at time $t=0$, $D_{max}$ is the capacity value of fractal dimension, i.e., the upper limit of fractal dimension, and $k$ is the initial growth rate of fractal dimension.

The formulae of speed and acceleration of fractal dimension increase can be derived from the logistic model. The fractal dimension growth rate can be defined by differentiation of fractal dimension with respect to time $t$, that is

$$S(t) = \frac{dD(t)}{dt} = kD(t)[1-\frac{D(t)}{D_{max}}], \qquad (2)$$

where $S(t)$ denotes the growth rate of fractal dimension of urban from. It is a measure for generalized speed. Further, taking the first derivative of equation (2) with respect to time $t$, or taking the second derivative of equation (1) yields the acceleration of fractal dimension increase, that is

$$a(t) = \frac{dS(t)}{dt} = \frac{d^2 D(t)}{dt^2} = k[1 - \frac{2D(t)}{D_{max}}]S(t), \tag{3}$$

where $a(t)$ refers to the acceleration of fractal dimension increase, indicating the acceleration of urban growth. Equation (3) indicates that, when $D(t)=D_{max}/2$, the acceleration is zero, namely, $a(t)=0$. When $D(t)<D_{max}/2$, we have $a(t)>0$, and when $D(t)>D_{max}/2$, we have $a(t)<0$.

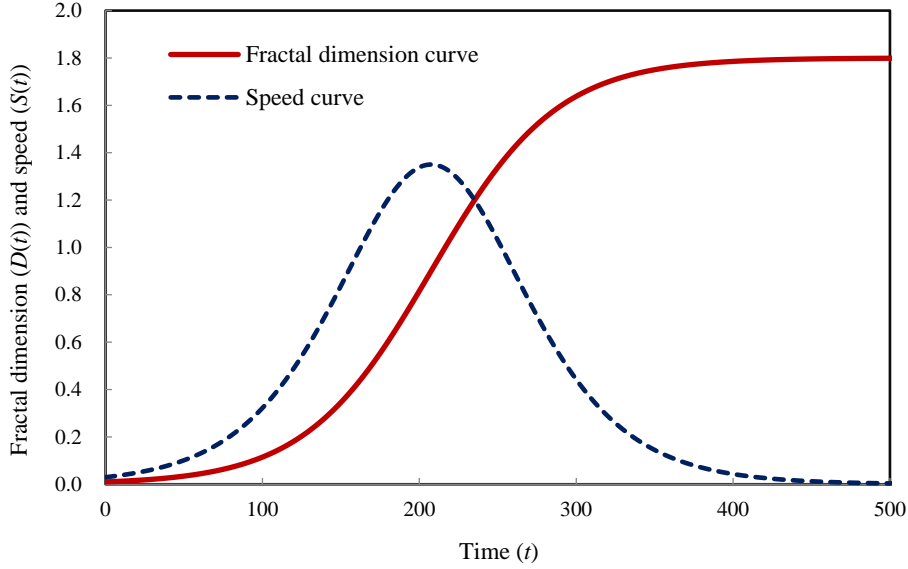

**Figure 1. Diagrammatic drawings of fractal dimension increase curves of urban form and corresponding fractal dimension increase speed curve** (**Note**: These are pure fractal dimension curves based on theoretical models. The parameter values as follows: $D_0=0.01$, $D_{max}=1.8$, $k=0.025$. The fractal dimension curve is drawn by equation (1), and the speed curve can be drawn by means of equation (2) or equation (5). In order to compare the fractal dimension curve and the fractal dimension speed curve in the same graph, the fractal dimension growth rate is magnified by 120 times.)

Further, three useful formulae of analysis and prediction of urban growth can be derived from the above equations. When the fractal dimension reaches half of its capacity value, the value of the fractal dimension increase velocity reaches its peak. Substituting equation (2) into equation (3) and then factoring it yields

$$a(t) = k^2 D(t)[1 - \frac{2D(t)}{D_{max}}][1 - \frac{D(t)}{D_{max}}], \tag{4}$$

which implies that, when $D(t)=0$, $D(t)=D_{max}/2$, and $D(t)=D_{max}$, the speed of fractal dimension increase equals 0. When $0<D(t)<D_{max}/2$, the acceleration is positive, i.e., $a(t)>0$; when $D_{max}/2<D(t)<D_{max}$, the acceleration is negative, i.e., $a(t)<0$. Inserting equation (1) into equation (2) yields

$$S(t) = \frac{kD_{max}(D_{max}/D_0 - 1)e^{-kt}}{[1+(D_{max}/D_0 - 1)e^{-kt}]^2}, \qquad (5)$$

which is the function of velocity over time. Substituting equation (1) into equation (4) yields

$$a(t) = k^2 \frac{D_{max}}{1+(\frac{D_{max}}{D_0}-1)e^{-kt}}[1-\frac{2}{1+(\frac{D_{max}}{D_0}-1)e^{-kt}}][1-\frac{1}{1+(\frac{D_{max}}{D_0}-1)e^{-kt}}], \qquad (6)$$

which is the function of acceleration over time. Using equation (5) and (6), we can draw the fractal dimension increase speed curve and acceleration curve (Figure 2).

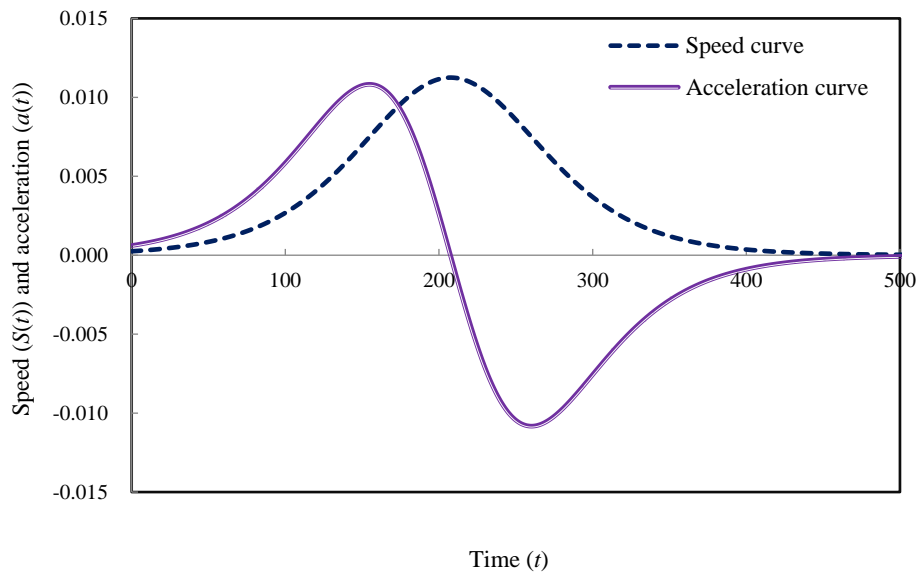

**Figure 2. Diagrammatic drawings of fractal dimension speed curves of urban form and corresponding fractal dimension acceleration curve** (**Note**: They are also pure theoretical curves. The parameter values are the same as those for Figure 1. The acceleration curve can be drawn by means of equation (4) or equation (6). In order to compare the fractal dimension speed curve and the fractal dimension acceleration curve in the same graph, the fractal dimension acceleration is magnified by 100 times.)

## 2.2 Stage division based on fractal dimension curves

Fractal dimension growth rate and acceleration curves can be used to define the dividing points of a process of urban growth. According to the principle of calculus, we can find the inflection point of fractal dimension growth by using the curve of growth rate (Figure 1); further, we can find the inflection points of growth rate curves by using the acceleration curve (Figure 2). The two sets of inflection points can be employed to identify the dividing points of urban phase transition. The derivative of $S(t)$ with respect to $D(t)$ is

$$\frac{dS(t)}{dD(t)} = k(1 - \frac{2D(t)}{D_{max}}) = 0. \tag{7}$$

According to the principle of extremum condition in higher mathematics, the peak value of fractal dimension increase speed is at

$$D_m(t) = \frac{D_{max}}{2} = \frac{\max[D(t)]}{2}, \tag{8}$$

which suggests the inflexion point of fractal dimension increase curve. When the fractal dimension $D(t)$ reaches half of the capacity value $D_{max}$, the speed of fractal dimension increase is the fastest, and then it will slow down. So, equation (8) corresponds to the peak of the fractal dimension increase speed curve in Figures 1 and 2. Substituting equation (8) into equation (2) yields

$$S_{max} = \max(\frac{dD(t)}{dt}) = \frac{kD_{max}}{2}(1 - \frac{1}{2}) = \frac{kD_{max}}{4}, \tag{9}$$

which gives the maximum rate of fractal dimension growth. As a special case, if $D_{max} = 2$ as given, then we have $S_{max} = k/2$.

Further, we can find the peak value and valley value of the acceleration of fractal dimension growth. Taking the derivative of equation (3) yields

$$\frac{da(t)}{dD(t)} = k^2[1 - \frac{6D(t)}{D_{max}} + \frac{6D(t)^2}{D_{max}^2}]. \tag{10}$$

If there are inflexion points on the curve of fractal dimension increase acceleration, then $da(t)/dL(t) = 0$, and thus we have

$$\frac{6D(t)^2}{D_{max}^2} - \frac{6D(t)}{D_{max}} + 1 = 0. \tag{11}$$

In practice, fractal dimension values can be normalized, and the lower value of fractal dimension is $D_{min} = 0$ in theory, thus the normalized value can be expressed as

$$D^*(t) = \frac{D(t)-D_{min}}{D_{max}-D_{min}} = \frac{D(t)}{D_{max}}. \tag{12}$$

Thus equation (11) can be simplified as a quadratic equation such as

$$6D^*(t)^2 - 6D^*(t) + 1 = 0. \tag{13}$$

Solving the quadratic function, equation (13), we can find two roots as follows

$$D^*(t) = \frac{D(t)}{D_{max}} = \frac{6 \mp \sqrt{6^2 - 4 \times 6 \times 1}}{2 \times 6} = \frac{3 \mp \sqrt{3}}{6}. \tag{14}$$

This suggests the two extremum values of fractal dimension increase acceleration, that is

$$D_l(t) = (\frac{1}{2} - \frac{1}{2\sqrt{3}})D_{max}, \tag{15}$$

$$D_u(t) = (\frac{1}{2} + \frac{1}{2\sqrt{3}})D_{max}. \tag{16}$$

In light of the mathematical reasoning, equations (15) and 16 corresponds to the valley and peak values of the fractal dimension increase acceleration curve in Figure 2, respectively. In fact, equations (15) and 16 can be derived by another way. Equation (4) can be expanded as follows

$$a(t) = k^2[D(t) - \frac{3D(t)^2}{D_{max}} + \frac{2D(t)^3}{D_{max}^2}]. \tag{17}$$

Derivative of equation (17) with respect to $t$ yields

$$\frac{da(t)}{dt} = k^2 \frac{dD(t)}{dt}[1 - \frac{6D(t)}{D_{max}} + \frac{6D(t)^2}{D_{max}^2}]. \tag{18}$$

Letting $da(t)/dt=0$ and $dD(t)/dt >0$ yields equation (11). The other steps are the same as what were shown above.

Now, a scheme of stage division of urban growth can be mathematically determined by means of the inflexion formulae. Equations (8), (15) and 16 give three threshold values of fractal dimension increase curve: lower limit, middle limit, and upper limit. Using the three threshold functions, we can divide the fractal dimension increase curve into four sections, which indicate four phases of urban growth (Figure 3). The first stage, $0<D(t)< D_{max}/2-D_{max}/(2\sqrt{3})$, can be termed initial stage of slow growth; the second stage, $D_{max}/2-D_{max}/(2\sqrt{3})< D(t)< D_{max}/2$, can be termed acceleration stage of fast growth; the third stage, $D_{max}/2<D(t)< D_{max}/2+D_{max}/(2\sqrt{3})$, can be termed deceleration stage of fast growth; and the fourth stage, $D_{max}/2< < D_{max}/2+D_{max}/(2\sqrt{3})<D(t)<D_{max}$, can be termed terminal stage of slow growth. Suppose that the a city evolves from a point ($D(t)=D_{min}=0$) into a fully filled region ($D(t)=D_{max}=d=2$). The three demarcation points are $D_l=d/2-d/(2\sqrt{3})=0.4226$, $D_m=d/2=1$, $D_u=d/2+d/(2\sqrt{3})=1.5774$.

**Table 1 Phase division results and inflexion points of urban growth based on logistic model of fractal dimension curves**

| Scheme (I) | Scheme (II) | Scheme (III) | Middle limit | Lower and upper limits |
|---|---|---|---|---|
| Acceleration | Initial slow growth | Initial slow growth | | $D_l(t)=(\frac{1}{2}-\frac{1}{2\sqrt{3}})D_{max}$ |
| | Celerity (fast growth) | Accelerated fast growth | $D_m(t)=\frac{D_{max}}{2}$ | |
| Deceleration | | Decelerated fast growth | | $D_u(t)=(\frac{1}{2}+\frac{1}{2\sqrt{3}})D_{max}$ |
| | Terminal slow growth | Terminal slow growth | | |

**Note**: $D_l(t)$ refers to the lower division, $D_m(t)$ to the middle division, and $D_u(t)$ to the upper division.

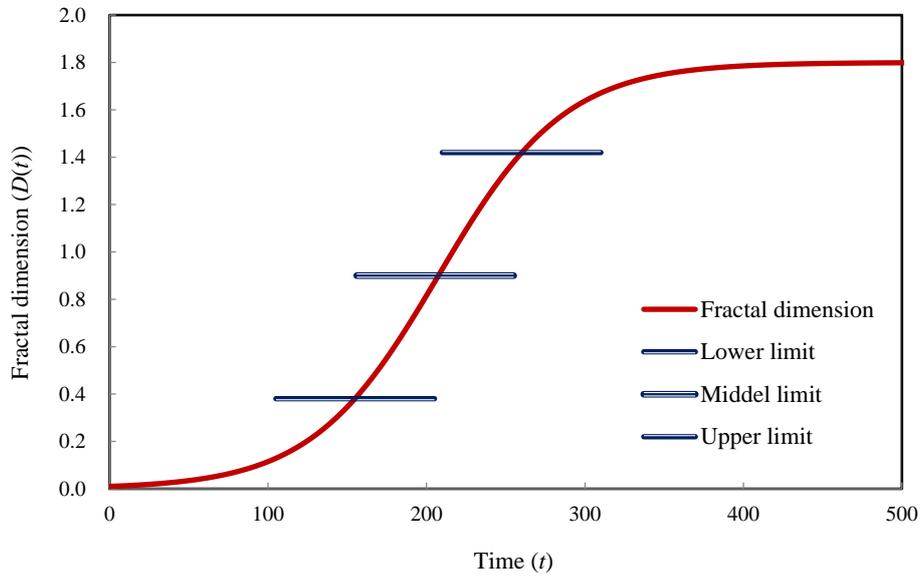

**Figure 3. Schematic diagram of stage division of urban growth based on logistic model of fractal dimension increase curves** (**Note**: The parameter values as follows: $D_0=0.01$, $D_{max}=1.8$, $k=0.025$. According to equations (8), (15) and (16), the threshold values are $D_l=0.3804$, $D_m=0.9$, $D_u=1.4196$, respectively.)

The results of stage division of urban growth based on fractal dimension of urban form can be simplified. Then we have three schemes as follows: (1) Four-stage scheme. The growing course can be divided into four stages: initial slow growth stage, accelerated fast growth stage, decelerated fast growth stage, and terminal slow growth stage. This is the basic scheme. (2) Three-stage scheme. The second stage and third stage can be combined into as a fast growth stage, which is termed celerity stage. Thus, the growing course can be divided into three stages: initial slow growth stage, celerity growth stage, and terminal slow growth stage. (3) Two-stage scheme. The first stage and the second stage can be merged into acceleration growth stage, and the third and fourth stage can be merged into the deceleration growth stage (Table 1).

**2.3 Stage division based on fractal dimension odds**

The basic meaning of fractal dimension is the degree of space filling. Suppose that the Euclidean dimension of the embedding space in which city fractals are defined is $d$. Thus the ratio of fractal dimension to the embedding dimension, $D/d$, indicates the probability of space filling in an urban region. Correspondingly, the difference, $1- D/d$, implies the probability of space remaining. In this case, under the condition that $D_{max}<d=2$, the fractal dimension odds can be defined as follows

$$Z(t) = \frac{D(t)}{d-D(t)} = \frac{D(t)}{2-D(t)} = \frac{D(t)/2}{1-D(t)/2}, \tag{19}$$

where $Z(t)$ denotes the fractal dimension odds at time $t$, $d=2$ refers to the Euclidean dimension of the embedding space of urban growing fractals. Equation (19) can be transformed as

$$D(t) = \frac{2Z(t)}{1+Z(t)}. \tag{20}$$

In theory, for the time $t=0$ and $t\to\infty$, equation (20) can be expressed as

$$D_0 = \frac{2Z_0}{1+Z_0}, \tag{21}$$

$$D_{max} = \frac{2Z_{max}}{1+Z_{max}}, \tag{22}$$

where $Z_0 = D_0/(2-D_0)$ and $Z_{max} = D_{max}/(1-D_{max})$ denote the initial value and capacity value of the fractal dimension odds, respectively. Substituting equation (1) into equation (19) yields

$$Z(t) = \frac{D_{max}}{2(1+(D_{max}/D_0-1)e^{-kt})-D_{max}}. \tag{23}$$

Further, inserting equations (21) and (22) into equation (23) yields

$$Z(t) = \frac{Z_{max}}{1+(Z_{max}/Z_0 -1)e^{-kt}}. \tag{24}$$

This suggests that, if a fractal dimension increase curve can be modeled with a three-parameter logistic function, the corresponding fractal dimension odds curve can also be modeled by a three-parameter logistic function. The second set of stage division scheme of urban growth process can be deduced by means of equation (24).

The derivation of the threshold functions of fractal dimension odds curve division is the same as that of the threshold functions of the fractal dimension increase curve. Taking the first derivative of equation (24) with respect to time $t$ yields the speed function as below

$$S^*(t) = \frac{dZ(t)}{dt} = kZ(t)[1-\frac{Z(t)}{Z_{max}}], \tag{25}$$

where $S^*(t)$ indicates the growth rate of fractal dimension odds at time $t$. Taking the first derivative of equation (25) or the second derivative of equation (24) yields the acceleration function as follows

$$a^*(t) = \frac{dS^*(t)}{dt} = \frac{d^2Z(t)}{dt^2} = k^2 Z(t)[1-\frac{2Z(t)}{Z_{max}}][1-\frac{Z(t)}{Z_{max}}], \tag{26}$$

where $a^*(t)$ indicates the acceleration of fractal dimension odds at time $t$. Equation (24) suggests three special values, $Z(t)=0$, $Z(t)=V_{max}/2$, $Z(t)=Z_{max}$, where the acceleration values are zero. Among the three values, the middle division points of the fractal dimension odds curve can be determined by

$$Z_m(t) = \frac{Z_{max}}{2}. \tag{27}$$

This is threshold function for the second dividing point. The derivative of $a^*(t)$ with respect to $Z(t)$ is a second order differential equation, that is

$$\frac{da^*(t)}{dZ(t)} = k^2[1-\frac{6Z(t)}{Z_{max}}+\frac{6Z(t)^2}{Z_{max}^2}]. \tag{28}$$

Assuming $da^*(t)/dV(t)=0$, we have a quadratic equation, for which two roots can be found as follows

$$V_1(t) = (\frac{1}{2}-\frac{1}{2\sqrt{3}})V_{max}, \tag{29}$$

$$V_u(t) = (\frac{1}{2} + \frac{1}{2\sqrt{3}})V_{max}. \tag{30}$$

These are threshold functions for the first and the third dividing points. Equations (27), (29), and (30) give the three demarcation points of the fractal dimension odds curve. Based on the three dividing points, the space filling process of urban growth can be divided into four stages (Table 2). The four stages can be simplified as three stages or two stages by stage merging.

Comparing the curves of fractal dimension increase speed and acceleration with those of the fractal dimension odds speed and acceleration shows a significant time difference. The peak values of fractal dimension odds increase speed and acceleration lag those of fractal dimension increase speed and acceleration. So are the valley values of the acceleration (Figure 4). This means that the fractal dimension increase curve is suitable for the phase division of a young city or the stage division of the whole growth process of a city, while the fractal dimension odds' increase curve is suitable for the stage division of metropolitan evolution or a mature city's growth. For many big cities, when we obtain the observation data, the growing peak has passed. In this case, the fractal dimension increase curve cannot reflect the whole picture of its growth process. However, the fractal dimension odds increase curve can reflect the peak of velocity and the fluctuation of acceleration.

**Table 2 Phase division results and inflexion points of urban space filling based on logistic model of fractal dimension odds curves**

| Scheme (I) | Scheme (II) | Scheme (III) | Middle limit | Lower and upper limits |
|---|---|---|---|---|
| **Acceleration** | Initial slow filling | Initial slow filling |  | $Z_1(t) = (\frac{1}{2} - \frac{1}{2\sqrt{3}})Z_{max}$ |
|  | Celerity (fast filling) | Accelerated fast filling | $Z_m(t) = \frac{Z_{max}}{2}$ |  |
| **Deceleration** |  | Decelerated fast filling |  | $Z_u(t) = (\frac{1}{2} + \frac{1}{2\sqrt{3}})Z_{max}$ |
|  | Terminal slow filling | Terminal slow filling |  |  |

**Note**: $Z_1(t)$ refers to the lower limit, $Z_m(t)$ to the middle limit, and $Z_u(t)$ to the upper limit.

The above-shown models based on fractal dimension odds are only suitable for the case of the fractal dimension capacity less than 2. If $D_{max}=2$, the fractal dimension odds will exponentially grow rather than logistically grow. This can be proved easily. In fact, if $D_{max} \to 2$, we have

$$Z_{max} = \lim_{D_{max} \to d} \frac{D(t)}{d - D(t)} \to \infty. \tag{31}$$

This indicates $1/Z_{max} \to 0$, and thus equation (24) will change to an exponential growth function as below

$$Z(t) = \frac{1}{1/Z_{max} + (1/Z_0 - 1/Z_{max})e^{-kt}} = Z_0 e^{kt}. \tag{32}$$

Based on equation (32), the growth speed and acceleration function are exponential functions. No inflexion point can be found in the curves of fractal dimension change speed and acceleration. Therefore, it is impossible to make stage division by means of speed and acceleration in this case.

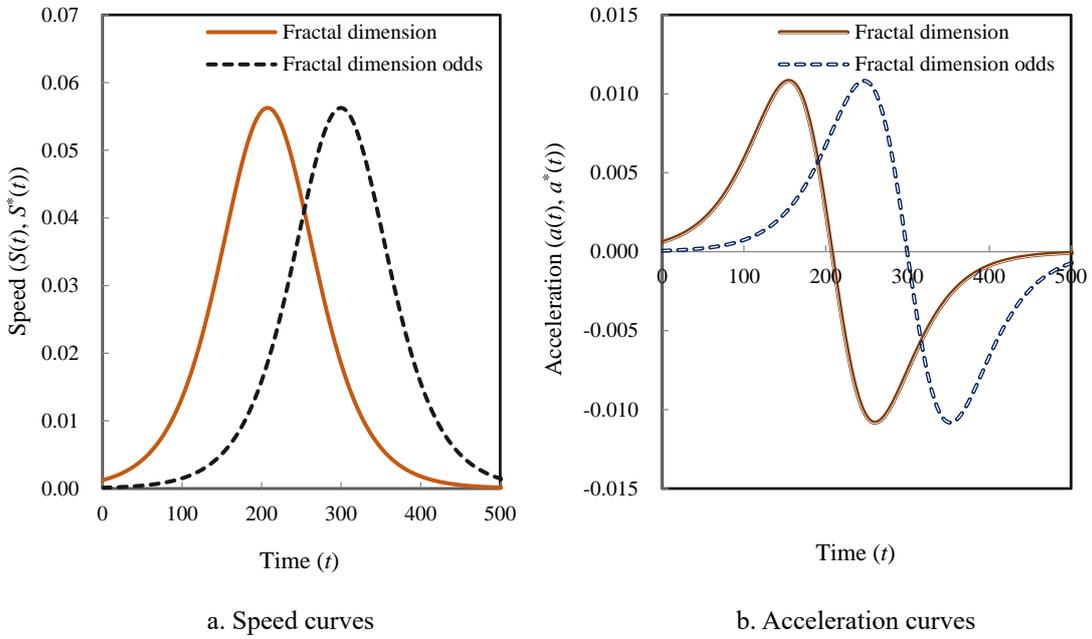

a. Speed curves  b. Acceleration curves

**Figure 4. Comparison between the speed-acceleration curves of fractal dimension increase and those of fractal dimension odds' increase** (Note: For facilitating intuitive comparison, the fractal dimension increase rate is magnified 5 times, and the fractal dimension odds increase acceleration is magnified 20 times. The time difference between two peaks or two valleys is about 92 years.)

## 3 Empirical analysis

### 3.1 Methods of stage division

A time series of fractal dimension of urban form is infinitely long in principle. This type of data series is suitable for theoretical analysis. In practical work, we always extract a period of data from the *time series* for dynamic analysis. The data sequence with finite length of time is termed *sample path* (Diebold, 2007). All empirical analyses are based on sample paths rather than real time series. The reliability of the stage division for urban growth depends on the accuracy of fractal dimension values, the length of sample path of fractal dimension parameters, and the growth mode of a city. In particular, the results of stage division rely to a degree on the definition of study area. Based on different size of study area, the stage division result for a city is different.

This section is mainly about the verification of theoretical models and illustration of analysis methods. The analytical process comprises the following steps.

**Step 1: calculate fractal dimension of urban form.** Many methods can be employed to estimate fractal dimension values of urban form (Batty and Longley, 1994; Frankhauser, 1998). The common approach is box-counting method (Benguigui *et al*, 2000; Shen, 2002). For the case studies in this paper, the fractal dimension have been estimated by other scholars.

**Step 2: build model of fractal dimension curves.** In the family of sigmoid functions, three logistic function can be used to describe fractal dimension curves of urban growth, including conventional logistic function, quadratic logistic function, and fractional logistic function (Chen, 2018). This research is based on the conventional logistic model of fractal dimension increase curves.

**Step 3: compute growth rate and acceleration of fractal dimension increase and draw curves for them.** Two sets of formulae can be utilized to make this calculation. The formulae for the growth rate and acceleration of fractal dimension are equations (2) and (4), which are based on differentials and represent continuous process of parameter computation. Equations (2) and (4) can be replaced by the difference formulae as follows

$$S(t) = \frac{\Delta D(t)}{\Delta t}, \qquad (33)$$

$$a(t) = \frac{\Delta S(t)}{\Delta t} = \frac{\Delta^2 D(t)}{\Delta t^2}, \qquad (34)$$

where $\Delta$ denotes difference operator. The difference formulae reflect discrete processes of parameter estimation. Accordingly, the differential-based formulae for the growth rate and acceleration of fractal dimension odds are equations (23) and (24), which can be substituted with the following difference formulae

$$S^*(t) = \frac{\Delta Z(t)}{\Delta t}, \tag{35}$$

$$a^*(t) = \frac{\Delta S^*(t)}{\Delta t} = \frac{\Delta^2 Z(t)}{\Delta t^2}. \tag{36}$$

The differential-based formulae based on continuous process are more accurate, while the difference-based formulae based on discrete process are simpler.

**Step 4: identify the peaks and valley values of growth rate and acceleration.** Two methods can be adopted to finish this task. One is formula method, the other is graphic method. The formula method is based on the formulae derived in Section 2. Using equations (8), (15) and (16), we can find the peak value of growth rate, the peak value and valley value of acceleration of fractal dimension growth. Using equations (27), (29) and (30), we can find the peak value of growth rate, the peak value and valley value of acceleration of fractal dimension odds growth. The graphic method is based on the curves of growth rate and acceleration. The peak value of growth rate and the peak value and valley value of acceleration can be found visually by using the graphs.

**Step 5: determine the schemes of stage division of urban growth.** The peak values and valley correspond to three dividing points of growing curves of fractal parameters.

Four cities are employed to make empirical analysis to show how to identify the growing stages of urban evolution. The cities are London of British, Baltimore of America, Tel Aviv of Israel, and Shenzhen of China. These cities were chosen because that the fractal dimension data of urban form in different years have been published (Batty and Longley, 1994; Benguigui *et al*, 2000; Shen, 2002; Man and Chen, 2020). The fractal dimension data of the four cities can be treated as sample path of fractal dimension time series. As indicated above, the methods of stage division of urban growth are based on sample paths of fractal parameters of cities. Every city has a specific history of development. With the help of mathematical models, we can only make moderate speculation about the past and future of a city based on the observed data. If the city encounters some special change in the process of urban development, the inference will be ineffective to some extent.

**3.2 Examples of urban stage division based on fractal dimension**

The first example is London, the capital of England. London is one of top-tier world cities (Knox and Marston, 2006). A set of fractal dimension of urban form were provided by Batty and Longley (1994). This set of fractal dimension values can be regarded as a sample path for stage division analysis. The advantage of this set of fractal dimension data is that it has a large time span, from 1820 to 1962. The disadvantages lie in two aspects: Firstly, there are fewer data points. Only eight values of fractal dimension are available. Secondly, the measurement caliber of the first four fractal dimension values may not be consistent with that of the last four ones. However, as a case of illustration for an analytical method, the results are worthy of reference. The sample path of fractal dimension of London's urban form can be modeled with the logistic function (Chen, 2018). The capacity parameter of logistic growth of fractal dimension is about $D_{max}$=1.8123 (Table 3). According to equation (8), the middle division, i.e., the second dividing point, corresponds to the fractal dimension value $D_m(t)= D_{max}/2 \approx 0.9062$; According to equations (15) and (16), the lower and upper divisions, i.e., the first and third dividing points, corresponds to the fractal dimension values $D_l(t)= 0.2113 D_{max} \approx 0.3830$, $D_u(t)= 0.7887 D_{max} \approx 1.4293$. By referring to the trend lines of fractal dimension growth, fractal dimension growth rate, and fractal dimension acceleration (Figure 5), the dividing years are around 1717, 1773, and 1829 (Table 4).

Table 3 Logistic models of fractal dimension curves and related materials of four cities: London, Baltimore, Tel Aviv, and Shenzhen

| City | Study area | Model | Goodness of fit | Data source |
|---|---|---|---|---|
| **London, UK** | Urban area | $\hat{D}(t) = \dfrac{1.8123}{1+0.3313e^{-0.0236t}}$ | 0.8568 | Batty and Longley, 1994 |
| **Baltimore, USA** | Urban area | $\hat{D}(t) = \dfrac{2}{1+1.7365e^{-0.0118t}}$ | 0.9658 | Shen, 2002 |
| **Tel Aviv, Israel** | Region 1 | $\hat{D}(t) = \dfrac{2}{1+0.2997e^{-0.0181t}}$ | 0.9622 | Benguigui et al, 2000 |
| | Region 2 | $\hat{D}(t) = \dfrac{1.9258}{1+0.3869e^{-0.0221t}}$ | 0.9907 | |
| | Region 3 | $\hat{D}(t) = \dfrac{1.8234}{1+0.3963e^{-0.0250t}}$ | 0.9886 | |
| **Shenzhen, China** | Metropolitan area | $\hat{D}(t) = \dfrac{1.7856}{1+0.5931e^{-0.1177t}}$ | 0.9912 | Man and Chen, 2021 |
| | Central area | $\hat{D}(t) = \dfrac{1.8869}{1+0.3121e^{-0.0823t}}$ | 0.9872 | |

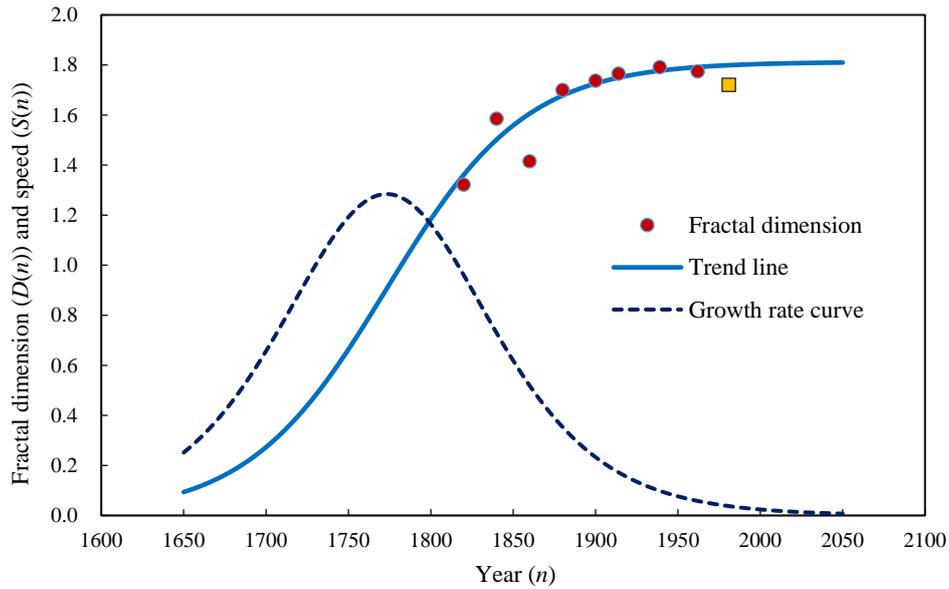

a. Fractal dimension and its growth rate

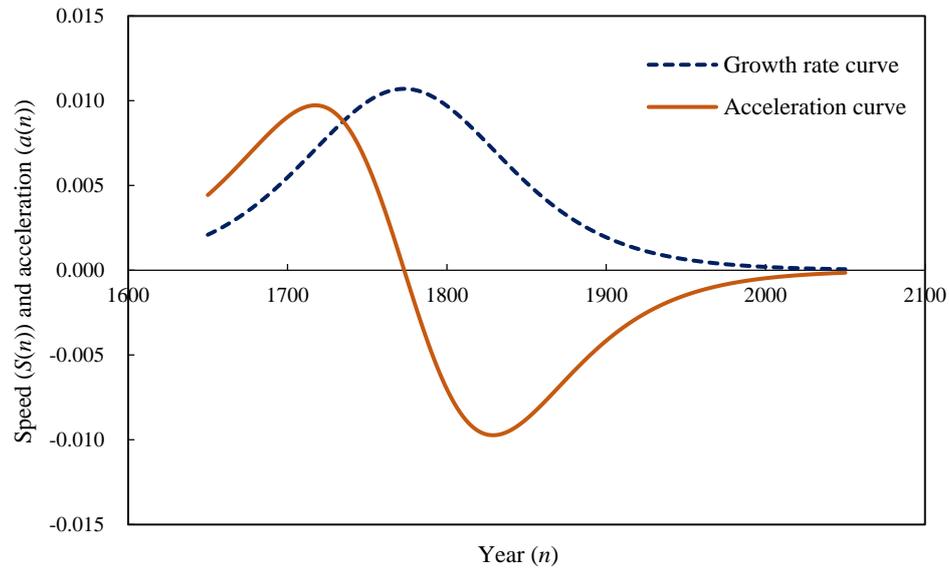

b. Fractal dimension growth speed and acceleration

**Figure 5. Fractal dimension, fractal dimension growth rate and acceleration of the urban form of London, UK** (**Note**: The fractal dimension values come from Batty and Longley (1994). The fractal dimension value in 1981 was calculated by Frankhauser (1994), who defined the study area differing from Batty and Longley (1994), and this data point is shown for reference only. For facilitating intuitive comparison, the fractal dimension growth rate is magnified 120 times in Figure 5(a), and the fractal dimension increase acceleration is magnified 100 times in Figure 5(b).)

The second example is Baltimore, a seaport city in northern Maryland of the United States of America (USA). It is also the most populous city in the state of Maryland. A set of fractal dimension of urban form were calculated with box-counting method by Shen (2002). The advantage of this set of fractal dimension data is that it has a larger time span, from 1792to 1992. The disadvantage is that there are only twelve available data points of fractal dimension. Baltimore's growing history is not as long as that of London. It can provide a case of stage division from different perspective. The sample path of fractal dimension of Baltimore's urban form can be modeled with the logistic function (Chen, 2012; Chen, 2018). The capacity parameter of logistic growth of fractal dimension is about $D_{max}$=2 (Table 3). According to equation (8), the middle division, i.e., the second dividing point, corresponds to the fractal dimension value $D_m(t)= D_{max}/2≈0.5$; According to equations (15) and (16), the lower and upper divisions, i.e., the first and third dividing points, corresponds to the fractal dimension values $D_l(t)= 0.2113D_{max}≈0.4226$, $D_u(t)= 0.7887D_{max}≈1.5774$. By referring to the trend lines of fractal dimension growth, fractal dimension growth rate, and fractal dimension acceleration (Figure 6), the dividing years are around 1727, 1839, and 1951 (Table 4). In fact, the development of Baltimore was after 1752. The first fractal dimension value used in this study is that for 1792. Using the logistic model of fractal dimension increase, we can predict the past fractal dimension of urban form for a city. However, the application of this prediction must be cautious. The special development period of a city often affect us to use logistic model to judge the past state.

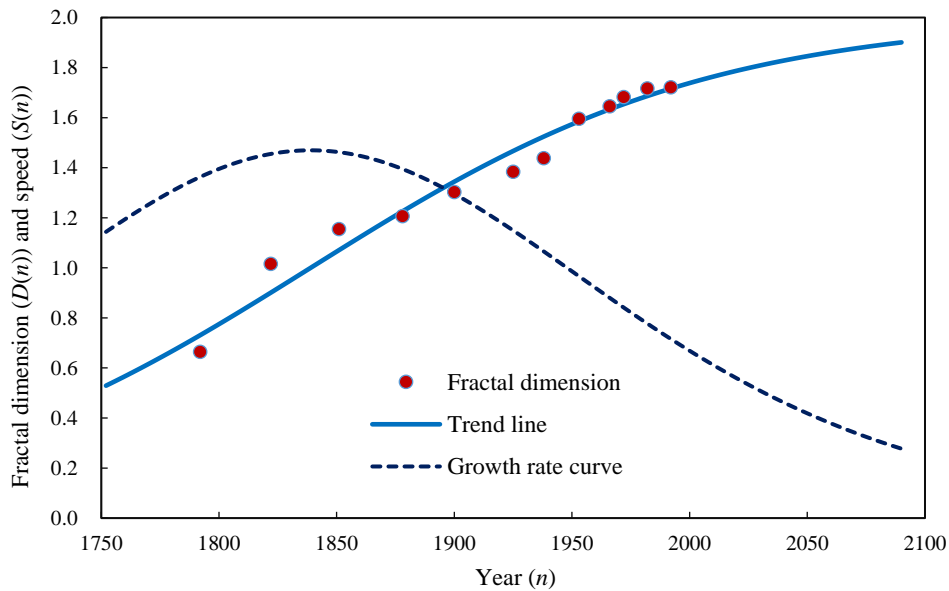

a. Fractal dimension and its growth rate

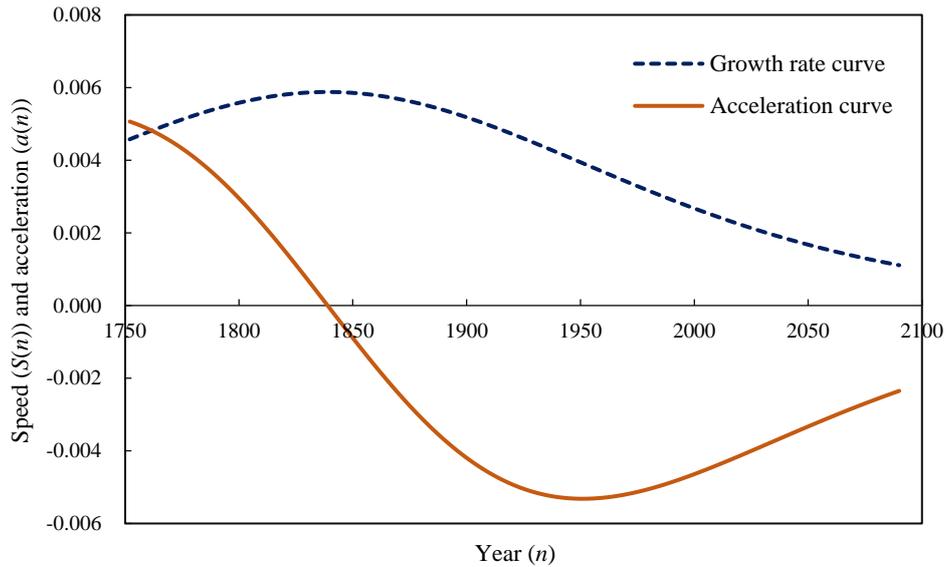

b. Fractal dimension growth rate and acceleration

**Figure 6. Fractal dimension, fractal dimension growth rate and acceleration of the urban form of Baltimore, USA** (**Note**: The fractal dimension values come from Shen (2002). For facilitating intuitive comparison, the fractal dimension growth rate is magnified 250 times in Figure 6(a), and the fractal dimension increase acceleration is magnified 200 times in Figure 6(b).)

The third example is Tel Aviv-Yafo, which is well known as Tel Aviv in international society. Tel Aviv is the most populous city in the Gush Dan metropolitan area of Israel. Three sets of fractal dimension were computed by Benguigui *et al* (2000). To research fractal evolution of Tel Aviv's urban morphology, Benguigui and his co-workers defined three study areas. The first region is the central part (region 1), the second region is the developed area, including the central part and the northeast part of the entire metropolis (region 2), and the third region is the entire metropolis (region 3). The development of these three regions was not synchronous. From the view of angle of fractal dimension increase, the central area was developed earlier than the entire ensemble, and the developed area was developed earlier than the whole metropolitan area. The progress differences of urban development can be seen from the peaks and valleys of fractal dimension growth rates and accelerations (Table 4). A similar problem to that of Baltimore is inevitably encountered for studying the growth stage of Tel Aviv. If we extrapolate the fractal dimension values beyond the sample data, the results sometimes depart the real trends. If we extrapolate back the fractal dimension values based on the present sample data to the past, the results sometimes depart the real trends significantly

(Figure 7).

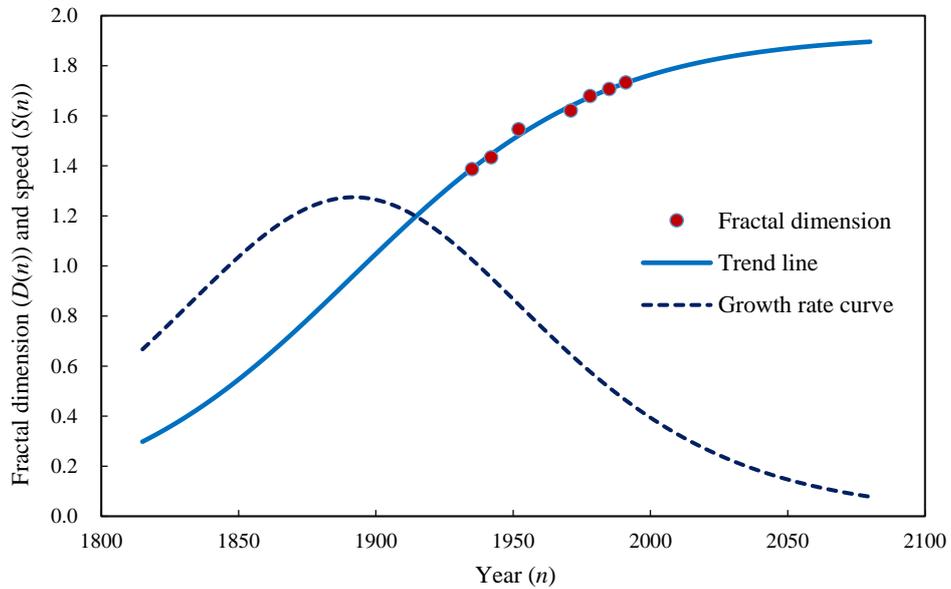

a. Fractal dimension and its growth rate

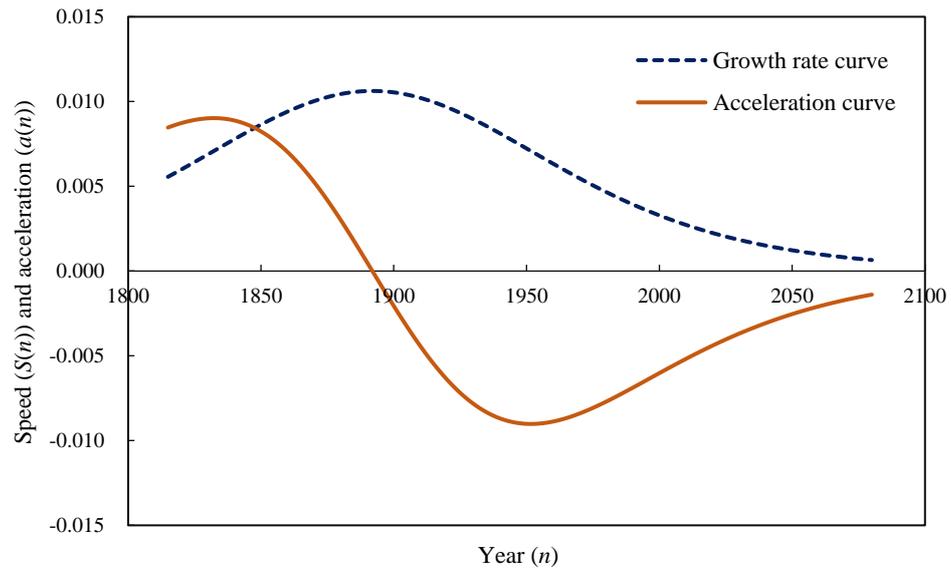

b. Fractal dimension growth rate and acceleration

**Figure 7. Fractal dimension, fractal dimension growth rate and acceleration of the urban form of Tel Aviv, Israel** (**Note**: The fractal dimension values come from Benguigui *et al* (2000). The study area is region 2. For facilitating intuitive comparison, the fractal dimension growth rate is magnified 120 times in Figure 7(a), and the fractal dimension increase acceleration is magnified 100 times in Figure 7(b).)

The fourth example is Shenzhen, an important city on the central coast of southern Guangdong province, People's Republic of China. Shenzhen is located on the east bank of the Pearl River estuary.

It is a model city after China's reform and opening up. In a sense, Shenzhen can be regarded as a shock city in the process of urbanization of China. A shock city is an embodiment of surprising and disturbing changes in economic, social, and cultural life (Knox and Marston, 2006). The fractal dimension was measured by Man and Chen (2020). The feature of this set of data is that the quality of remote sensing images used for fractal dimension measurement is very good and the fractal dimension values are relatively reliable. The disadvantage is that the time span of sample path is short and there are only 12 data points involving 32 years. Due to the short sample path and special developing history, we can only identify two growth stages of Shenzhen. Shenzhen was originally a small town in southern China. In 1980, Shenzhen was established as China's first special economic zone. The central government of China does its best to support the development of Shenzhen. In 1982, the growth rate of fractal dimension of Shenzhen metropolitan area reached its peak. In 1993, the acceleration of fractal dimension growth fell into the valley (Table 4). If we focus on the central area of Shenzhen city, we can only find one dividing points, that is, the valley value of acceleration of fractal dimension growth.

**Table 4 The dividing points of four growing stages of urban growth based on the logistic model of fractal dimension curves**

| Type | Dividing point | London | Baltimore | Tel Aviv (1) | Tel Aviv (2) | Tel Aviv (3) | Shenzhen (MA) | Shenzhen (CA) |
|---|---|---|---|---|---|---|---|---|
| **Lower division** | $D_l$ | 0.3830 | 0.4226 | 0.4226 | 0.4070 | 0.3853 | 0.3773 | 0.3987 |
| | Year | 1717 | (1727) | 1793 | 1832 | 1845 | (1971) | (1956) |
| **Middle division** | $D_m$ | 0.9062 | 1 | 1 | 0.9629 | 0.9117 | 0.8928 | 0.9435 |
| | Year | 1773 | 1839 | 1864 | 1892 | 1898 | 1982 | (1972) |
| **Upper division** | $D_u$ | 1.4293 | 1.5774 | 1.5774 | 1.5188 | 1.4381 | 1.4083 | 1.4882 |
| | Year | 1829 | 1951 | 1932 | 1952 | 1951 | 1993 | 1988 |

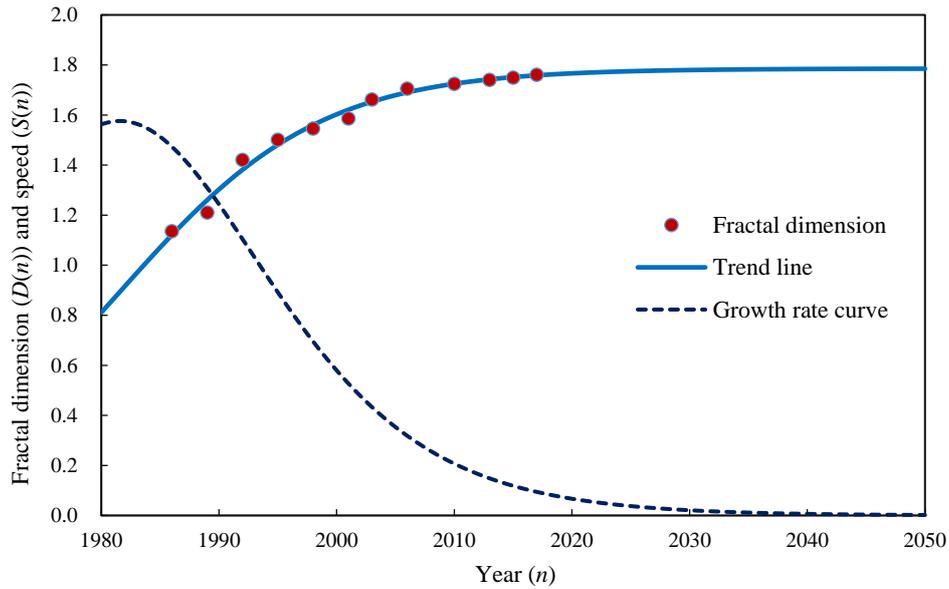

a. Fractal dimension and its growth speed

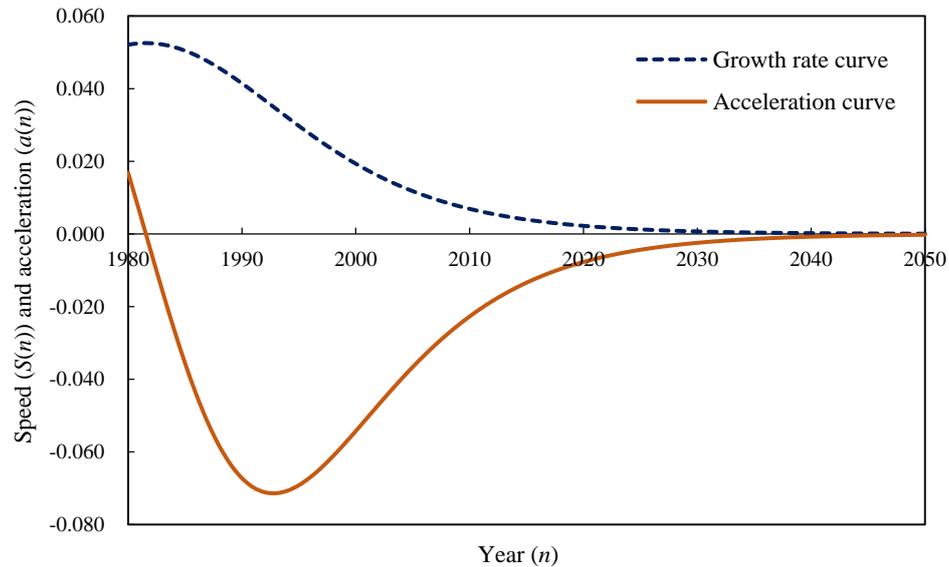

b. Fractal dimension growth rate and acceleration

**Figure 8. Fractal dimension, fractal dimension growth rate and acceleration of the urban form of Shenzhen, China** (**Note**: The fractal dimension values come from Man and Chen (2020). The study area is metropolitan area. For facilitating intuitive comparison, the fractal dimension growth rate is magnified 30 times in Figure 8(a), and the fractal dimension increase acceleration is magnified 30 times in Figure 8(b).)

### 3.3 Examples of urban stage division based on fractal dimension odds

Fractal dimension denotes the absolute degree of space filing, while fractal dimension odds implies the relative degree of space filling. The fractal dimension odds can reflects urban growth and its stage characteristics from another perspective. If the fractal dimension capacity $D_{max}=2$, the curve of fractal dimension odds will satisfy exponential function and show no stage properties. If the fractal dimension capacity $D_{max}<2$, the curve of fractal dimension odds will satisfy logistic function and can be divided into four stages. According to the mathematical models shown above, the fractal dimension capacity parameters of Baltimore and region 1 of Tel Aviv can be treated as 2, and thus cannot be divided into several stages. The curves of fractal dimension odds of London, Shenzhen, and Regions 2 and 3 of Tel Aviv can be described with logistic function (Table 5).

Table 5 Logistic models of fractal dimension odds curves and related materials of four cities: London, Baltimore, Tel Aviv, and Shenzhen

| City | Study area | Model | Goodness of fit | Data source |
|---|---|---|---|---|
| London, UK | Urban area | $\hat{Z}(t) = \dfrac{9.6553}{1+3.5305e^{-0.0236t}}$ | 0.8568 | Batty and Longley, 1994 |
| Baltimore, USA | Urban area | $\hat{Z}(t) = 0.5759e^{-0.0118t}$ | 0.9658 | Shen, 2002 |
| Tel Aviv, Israel | Region 1 | $\hat{Z}(t) = 3.3362e^{-0.0181t}$ | 0.9622 | Benguigui *et al*, 2000 |
| | Region 2 | $\hat{Z}(t) = \dfrac{25.9542}{1+10.4283e^{-0.0221t}}$ | 0.9907 | |
| | Region 3 | $\hat{Z}(t) = \dfrac{10.3250}{1+4.4878e^{-0.0250t}}$ | 0.9886 | |
| Shenzhen, China | Metropolitan area | $\hat{Z}(t) = \dfrac{8.3284}{1+5.5323e^{-0.1177t}}$ | 0.9912 | Man and Chen, 2021 |
| | Central area | $\hat{Z}(t) = \dfrac{16.6835}{1+5.5195e^{-0.0823t}}$ | 0.9872 | |

Table 6 The dividing points of four growing stages of urban growth based on the logistic model of fractal dimension odds curves

| Type | Point | London | Baltimore | Tel Aviv (1) | Tel Aviv (2) | Tel Aviv (3) | Shenzhen (MA) | Shenzhen (CA) |
|---|---|---|---|---|---|---|---|---|
| Lower division | $Z_l$ | 2.0404 | -- | -- | 5.4848 | 2.1819 | 1.7600 | 3.5256 |
| | Year | 1818 | -- | -- | 1982 | 1942 | 1989 | 1991 |
| Middle division | $Z_m$ | 4.8277 | -- | -- | 12.9771 | 5.1625 | 4.1642 | 8.3417 |
| | Year | 1873 | -- | -- | 2041 | 1995 | 2001 | 2007 |
| Upper division | $Z_u$ | 7.6149 | -- | -- | 20.4694 | 8.1431 | 6.5684 | 13.1578 |
| | Year | 1929 | -- | -- | 2101 | 2048 | 2012 | 2023 |

The curves of fractal dimension odds of London can be divided into four stages. The dividing points are $0.2113Z_{max}=2.0404$, $Z_{max}/2=4.8277$, and $0.7887Z_{max}=7.6149$, respectively, and the corresponding years are 1818, 1873, and 1929, respectively (Figure 9, Table 6).

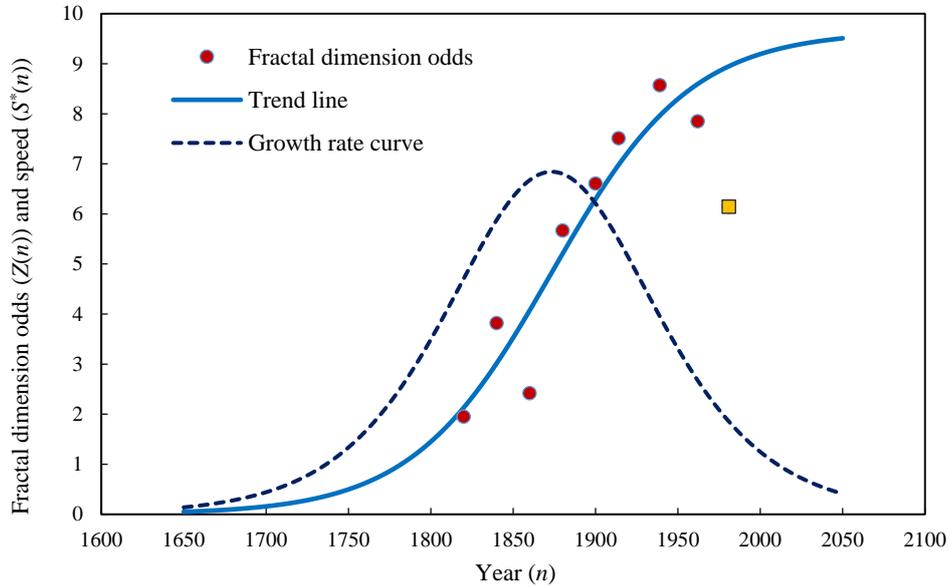

a. Fractal dimension odds and its growth rate

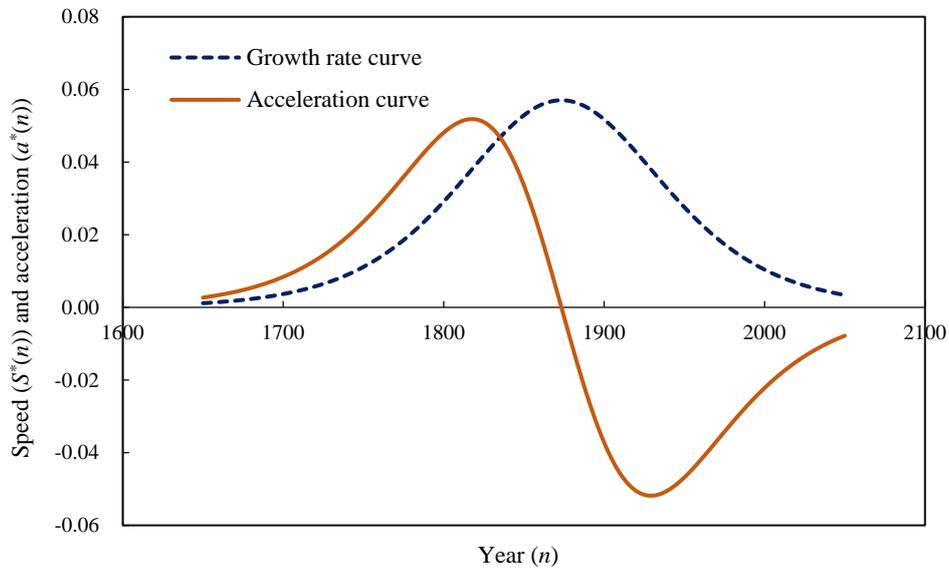

b. Fractal dimension odds growth speed and acceleration

**Figure 9. Fractal dimension odds, growth rate and acceleration of fractal dimension odds of the urban form of London, UK** (**Note**: The fractal dimension odds values were turned by the fractal dimension values come from Batty and Longley (1994) and Frankhauser (1994). The value in 1981 is treated as an outlier due to the data caliber. For facilitating intuitive comparison, the fractal dimension odds growth rate is magnified 120 times in Figure 9(a), and the fractal dimension odds increase acceleration is magnified 100 times in Figure 9(b).)

The curves of fractal dimension odds of Regions 2 and 2 of Tel Aviv can be modeled with logistic function. Four region 2, the dividing points are $0.2113Z_{max} = 5.4848$, $Z_{max}/2=12.9771$, and $0.7887Z_{max}=20.4694$, respectively, and the corresponding years are 1982, 2041, and 2101, respectively (Figure 10). Four region 3, the dividing points are $0.2113Z_{max} = 2.1819$, $Z_{max}/2=5.1625$, and $0.7887Z_{max}=8.1431$, respectively, and the corresponding years are 1942, 1995, and 2048, respectively (Table 6). The peak values and valley value of region 2 come later than those of region 3.

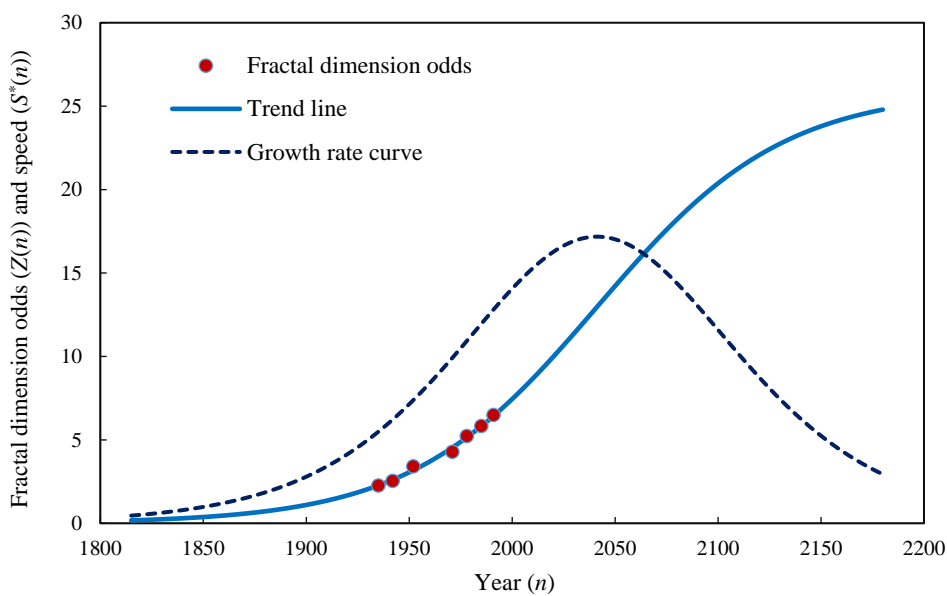

a. Fractal dimension odds and its growth speed

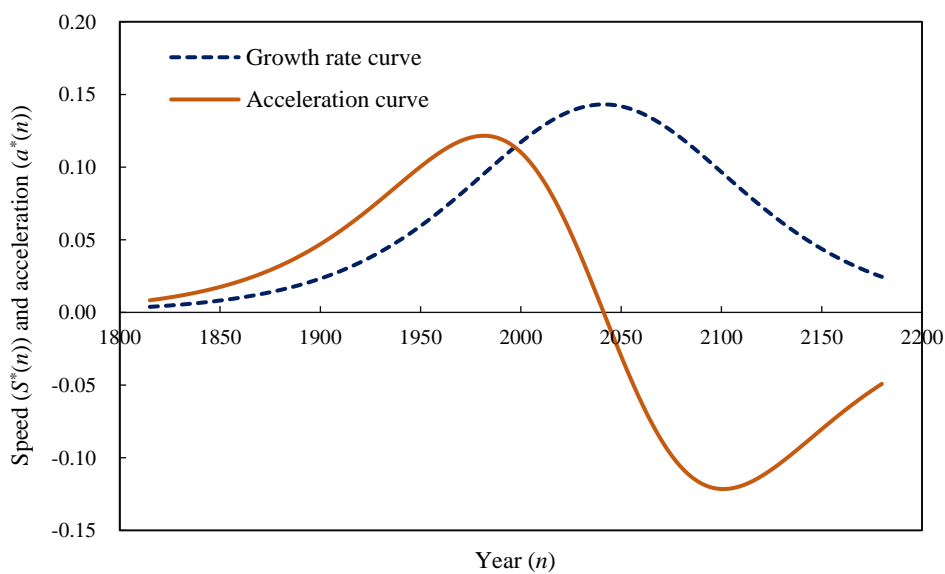

b. Fractal dimension odds growth speed and acceleration

**Figure 10. Fractal dimension odds, growth rate and acceleration of fractal dimension odds of the urban form of Tel Aviv, Israel** (**Note**: The fractal dimension odds values were turned by the fractal dimension values come from Benguigui *et al* (2000). The study area is region 2. For facilitating intuitive comparison, the fractal dimension odds growth rate is magnified 120 times in Figure 9(a), and the fractal dimension odds increase acceleration is magnified 100 times in Figure 9(b).)

The curves of fractal dimension odds of Shenzhen can be modeled with logistic function and thus can be divided into four stages. Four metropolitan area, the dividing points are $0.2113Z_{max} = 1.7600$, $Z_{max}/2 = 4.1642$, and $0.7887Z_{max} = 6.5684$, respectively, and the corresponding years are 1989, 2001, and 2012, respectively (Figure 11). Four central area, the dividing points are $0.2113Z_{max} = 3.5256$, $Z_{max}/2 = 8.3417$, and $0.7887Z_{max} = 13.1578$, respectively, and the corresponding years are 1991, 2007, and 2023, respectively (Table 6). The peak values and valley value of central area come later than those of metropolitan area.

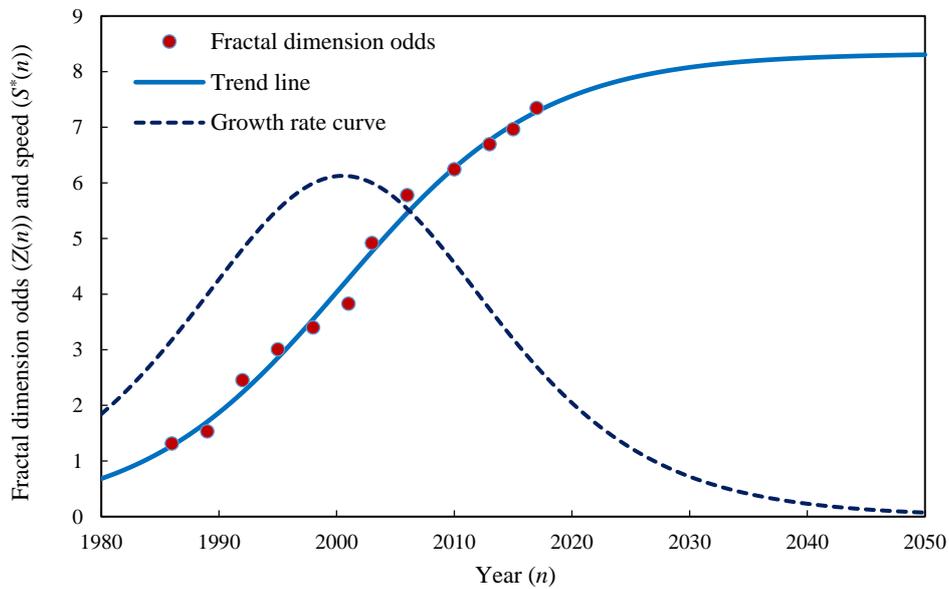

a. Fractal dimension odds and its growth rate

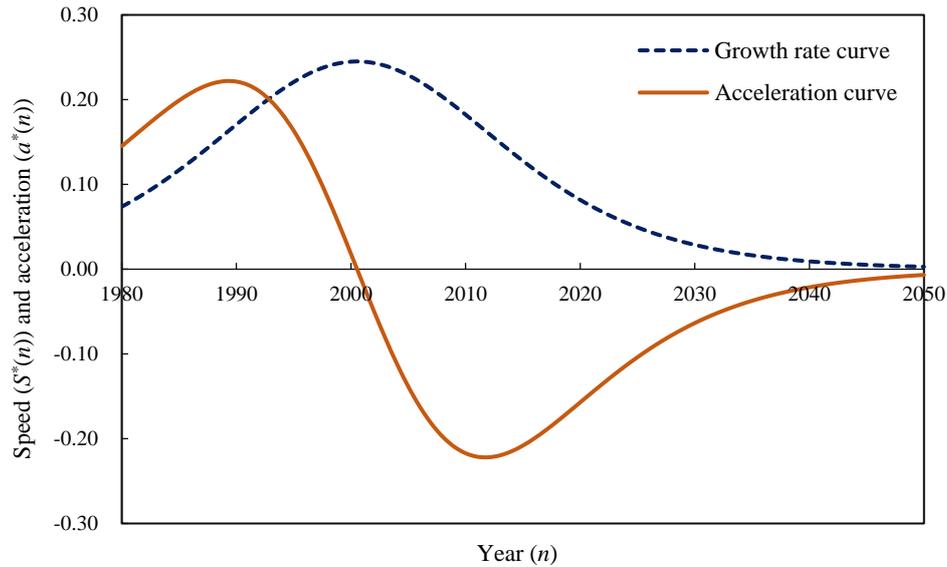

b. Fractal dimension odds growth rate and acceleration

**Figure 11. Fractal dimension odds, fractal dimension odds growth rate and acceleration of the urban form of Shenzhen, Chen** (**Note**: The fractal dimension odds values were turned by the fractal dimension values come from Man and Chen (2020). The study area is Metropolitan area. For facilitating intuitive comparison, the fractal dimension odds growth rate is magnified 25 times in Figure 11(a), and the fractal dimension odds increase acceleration is magnified 20 times in Figure 11(b).)

**3.4 Stage division of urban growth based on population size**

For reference and comparison, it is necessary to examine the growth curves of urban population and the corresponding stage division results. Two central variables in the studies on spatial dynamics of urban evolution are population and wealth (Dendrinos, 1992). Population may represent the first order dynamics of urban growth (Arbesman, 2012). Urban evolution is the process of interplay between human activities and land use. The fractal dimension is the spatial index of urban form reflecting land use. It is hard to measure the fractal dimension of urban population. However, we can model the growing course of urban population. One example is British city, London. From 1801 to 1951, the population change of London take on logistic curve. After 1951, the population of Greater London declined and began to rise after 1981. The peak value of London's population growth rate appeared at about 1882, the peak and valley values of London's population growth acceleration appeared at about 1837 and 1928. The time interval is about 45 to 46 years (Appendix 1). Another example is the American city, Baltimore. The city experienced a process of population growth first and then decline. From 1775 to 1960, Baltimore's population change can be modeled with logistic function. After 1960, the population of Baltimore began to decrease and deviated from the trend line of logistic growth. The peak value of Baltimore's population growth rate appeared at about 1893, the peak and valley values of Baltimore's population growth acceleration appeared at about 1860 and 1925. The time interval is about 32 to 33 years (Appendix 2).

Fractal dimension increase of urban form reflects the growing process of urban land use measured by impervious area. Population growth differs from land use growth. From the population growth curves of London and Baltimore, we can see the following commonalities. **First, population and land use growth are not synchronized.** The peak of population growth rate is significantly later than that of fractal dimension growth rate. **Second, urban population growth is faster than land use growth.** The time difference between peak and valley of the acceleration curve of population growth is significantly smaller than that of the acceleration curve of fractal dimension growth. **Third, land use growth is much more stable than population growth.** Even if the population size of a city begins to decline, the fractal dimension of urban form does not decline accordingly.

## 4 Discussion

The empirical analyses support the theoretical results of mathematical derivation for urban phase transition. The process of urban growth can be divided into two, three, or four stages. The basic model is based on the four-stage division pattern. However, different fractal parameters give different schemes. The time series of fractal dimension can be applied to the stage division of urban growth in a complete growing process, and time series of the fractal dimension odds is suitable for the stage division of a mature city or large city. The stages based on fractal dimension reflect the process of urban growth, while the stages based on fractal dimension odds reflect the process of urban space filling. The former mainly reflects the denotative development of a city (the expansion of urban space), while the latter mainly reflects the connotative development of the city (the optimization of urban space). In fact, for urban land use form, the fractal dimension can be approximately expressed as

$$D = \frac{2\ln A}{\ln A_e}, \tag{37}$$

where $A$ denotes the impervious area, and $A_e$ refers to the area of total urban region. Accordingly, the fractal dimension odds can be given by

$$Z = \frac{\ln A}{\ln A_e - \ln A}. \tag{38}$$

In fact, equation (38) can be derived from equation (37) by means of equation (19). Through equations (37) and (38), we can understand the geographical meaning of fractal dimension and fractal dimension odds of urban form intuitively.

    The division of urban growth stage cannot get absolutely definite result. The result of phase division depends on the definition of study area. The larger the study area is, the later the peaks and valley of urban growth rate and acceleration come. On the other hand, the accuracy of urban growth stage division depends on the accuracy of fractal dimension values and the length of fractal dimension sample path. The main shortcomings of this research are as follows. First, data quality. The sample paths of the studied cities are short due to lack of fractal dimension data of sample cities in early years. The accurate maps and the remote sensing images of a city appears later, and thus we cannot obtain the fractal dimension values in earlier years. Although the trend line can be extended to the early stage with the help of logistic model, the reliability of the predicted values for the historical data is relatively low. Second, model limitations. The fractal dimension increase processes of urban form can be modeled by at least three types of functions: common logistic function, quadratic logistic function, and fractional logistic function. However, in this work, only the common logistic model was taken into account. Other types of sigmoid functions such as quadratic logistic function and fractional logistic function have not been investigated for the time being. Third, absence of explanation. The stage division of urban growth is mainly a description model. To explain the formation of urban phrase transition patterns, we need logistic models of fractal dimension increase. In equation (1), the variable of time, $t$, is actually a dummy variable, which is time dummy (Diebold, 2007). If there is no real causal relationship between an independent variable and a dependent variable, then the independent variable belongs to a dummy variable. Dummy variables include time dummy, distance dummy, and categorical variables (nominal variable, indicator variables). Suppose that the real explanatory variables behind time dummy is $x_j$ ($j$=1, 2,…, $m$). In the simplest case, we have a linear decomposition as follows

$$t = \frac{1}{k}(a + b_1 x_1 + b_2 x_2 + \cdots + b_m x_m), \tag{39}$$

where *m* denotes the number of influence factors. Substituting equation (39) into equation (1) yields a logistic regression model based on fractal dimension ratio as below

$$\frac{D(t)}{D_{max}} = \frac{1}{1 + e^{-a - b_1 x_1 - b_2 x_2 - \cdots - b_m x_m}}, \tag{40}$$

where *a* and $b_j$ refers to logistic regression coefficients. Based on equation (19), we can derive a logit transform of fractal dimension odds as follows

$$\ln \frac{D(t)/D_{max}}{1 - D(t)/D_{max}} = a + \sum_{j=1}^{m} b_j x_j. \tag{41}$$

Using equations (39) to (41), we can explain the stage division of urban growth by the logistic regression analysis based on fractal parameters and related observed data.

The studies similar to this work have not been found in the previous literature so far. The novelty of this work lies in two set of models of stage division of urban growth based on the logistic model of fractal dimension curves. The stage division can provide new approach of understanding urban evolution. Two significant findings are worth emphasizing here. *Firstly, compared with the whole urban area, the dividing points of the fractal dimension curve of the central area of a city appear earlier, but the dividing points of fractal dimension odds come later* (Compare Table 4 with Table 6). This means that the initial development of the city begins in the central area, and the final stage also concludes in the central area. *Secondly, the change of spatial pattern reflected by fractal dimension values of urban land use is not consistent with the process of population growth.* The peak of urban population growth rate is significantly behind the peak of fractal dimension growth of urban morphology. However, when the urban population has stopped growing or even begin decreasing, the fractal dimension value of urban form continues to increase slowly. This suggests that fractal dimension is a stable parameter to describe urban growth.

Next, we can promote and develop the stage division models of urban growth based on fractal parameters in three directions. First, the models can be generalized to spatial entropy. On the one hand, fractal dimension is actually defined on the basis of entropy. On the other hand, normalized fractal dimension proved to equal normalized spatial entropy (Chen, 2020a). The spatial entropy curves can be modeled with logistic function (Chen, 2020b). Using spatial entropy, we can make stage division for urban growth. Second, the models can be generalized to general sigmoid functions. In fact, fractal dimension curves of urban growth can be modeled by logistic function, quadratic logistic function, and fractional logistic function. Different functions are suitable for different types of cities. The ideas of phase division based on growth rate and acceleration can be generalized to the quadratic logistic model and fractional logistic models. The processes and results of mathematical derivation are more complicated. For the quadratic logistic function, the curve is not symmetric, and the trend line cannot be extended to the past. Third, the models can be generalized to multifractal measures. For a monofractal system, one fractal parameter is enough to characterize its spatial structure. In contrast, for a multifractal system, we need several sets of fractal parameters to describe it. At least, we can make stage division of urban growth based on capacity dimension, information dimension, and correlation dimension.

# 5 Conclusions

The time series of fractal dimension values of cities can be employed to identify the developing phases of urban growth. The basis rests with that the fractal dimension increase curves can be described with sigmoid functions. Although we have provided several research cases, we cannot guarantee that the results of stage division are all consistent with the actual situations of urban growth. The main factors of influencing the stage division of urban growth include the definition of study area, the algorithms for fractal dimension estimation, the accuracy of fractal dimension measurement, the length of sample path of fractal dimension values. This paper is focused on theoretical method of stage division based on the logistic process of urban growth. Suppose that the fractal dimension curve of a city growth can be model with the logistic function. The main conclusions can be reached as follows. **First, based on fractal dimension values, the growing course of a city can be divided into four stages: initial slow growth stage, accelerated fast grow stage, decelerated fast growth stage, terminal slow growth stage.** The dividing points of the fractal dimension curve are $0.2113D_{max}$, $0.5D_{max}$, $0.7887D_{max}$, where $D_{max}$ denotes the capacity value of the fractal dimension of urban form. This method can be applied to young cities or a complete developing process of a large city (from $D(t) < 0.2113D_{max}$ to $D(t) > 0.7887D_{max}$). The precondition of effective stage division of urban growth is that the sample path of fractal dimension values is long enough and the fractal dimension values are exact enough. **Second, based on fractal dimension odds values, the space filling course of a city can also be divided into four stages: initial slow filling stage, accelerated fast filling stage, decelerated fast filling stage, terminal slow filling stage.** The dividing points of the fractal dimension curve are $0.2113Z_{max}$, $0.5Z_{max}$, $0.7887Z_{max}$, where $Z_{max} = D_{max}/(2-D_{max})$ denotes the capacity value of the fractal dimension of urban form. This method is suitable for the mature cities or the deceleration growth stage of a large city ($D(t) > 0.5D_{max}$). The precondition of effective stage division of urban space filling is that the fractal dimension values are exact enough and the capacity value of fractal dimension is less than the Euclidean dimension of embedding space ($D_{max} < 2$). **Third, the locations of the dividing points depend on the definition of the study area of a city.** This is easy to understand. Whether we focus on urban growth or urban space filling, the development speeds will rely on the size of urban space definition. The growth rate of a city relative to large region is slower than that relative to small region. The larger the study area we defined, the slower the growth of the city. Accordingly, the dividing points will move back. In practice, the study area of a city can be defined by the research objectives, the status of city development, and comparability between different cities.

**Acknowledgement:**

This research was sponsored by the National Natural Science Foundations of China (Grant No. 41671167). The support is gratefully acknowledged.

# Appendix 1: Stage division of London's urban growth based on population size

The logistic models of population growth of London can be built by observational data. Based on the sample path from 1801 to 1951, the model of Greater London's population growth is as below:

$$\hat{P}(t) = \frac{9{,}647{,}810}{1+10.3784e^{-0.0288t}} \ .$$

The goodness of fit is about $R^2 = 0.9850$ (Figure A1).

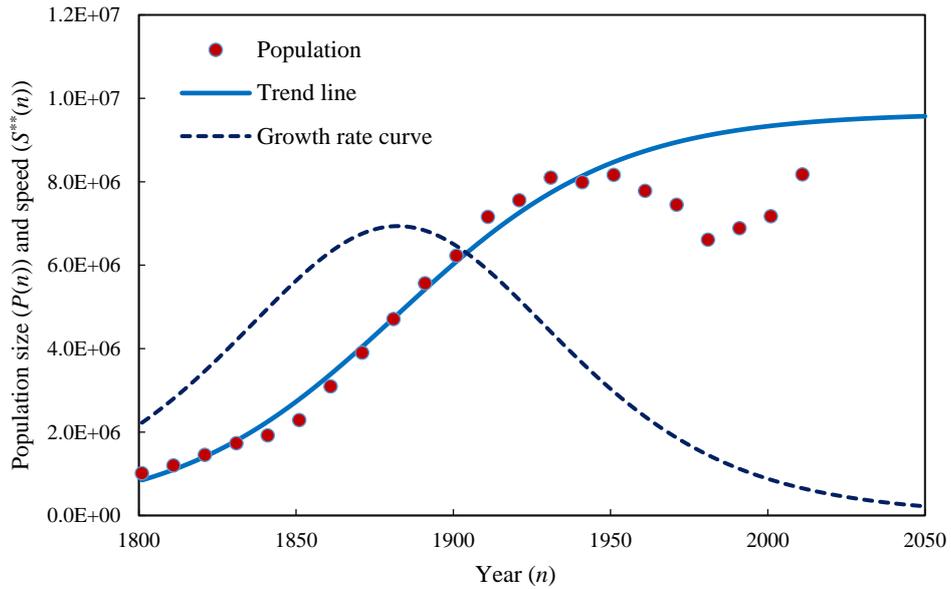

a. Population size and its growth speed

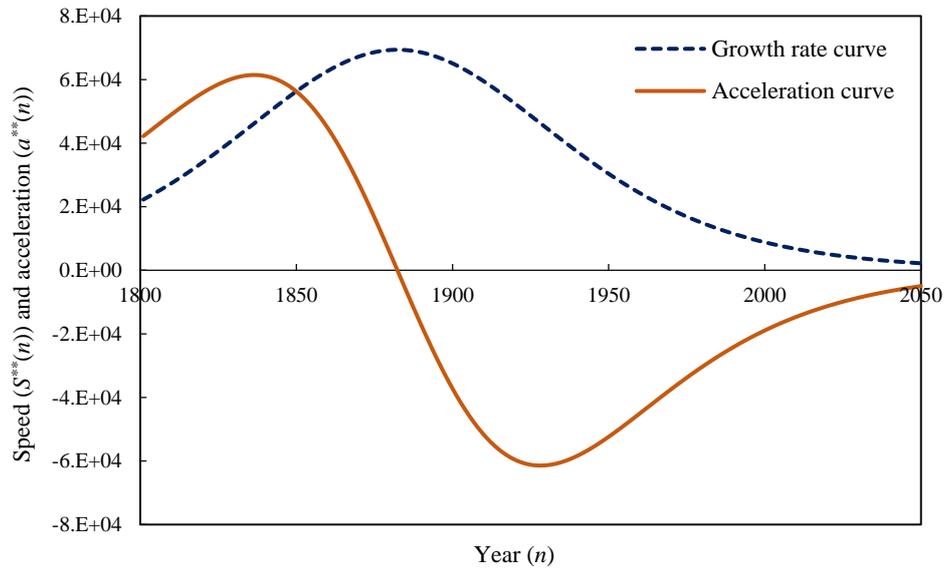

b. Population growth rate and acceleration

**Figure A1. Population size, population growth rate and acceleration of Greater London, UK** (Note: Population data come from http://en.wikipedia.org/wiki/Demography_of_London. For facilitating intuitive comparison, the population growth rate is magnified 100 times in Figure A1 (a), and the population growth acceleration is magnified 80 times in Figure A1 (b).)

# Appendix 2: Stage division of Baltimore's urban growth based on population size

Using the similar way, we can obtain the logistic model of Baltimore's population growth. Based on the sample path of population size from 1775 to 1960, Baltimore's urban growth can be modeled as follows

$$\hat{P}(t) = \frac{1,000,000}{1+301.8407e^{-0.0406t}} .$$

The goodness of fit is about $R^2$=0.9895 (Figure A2).

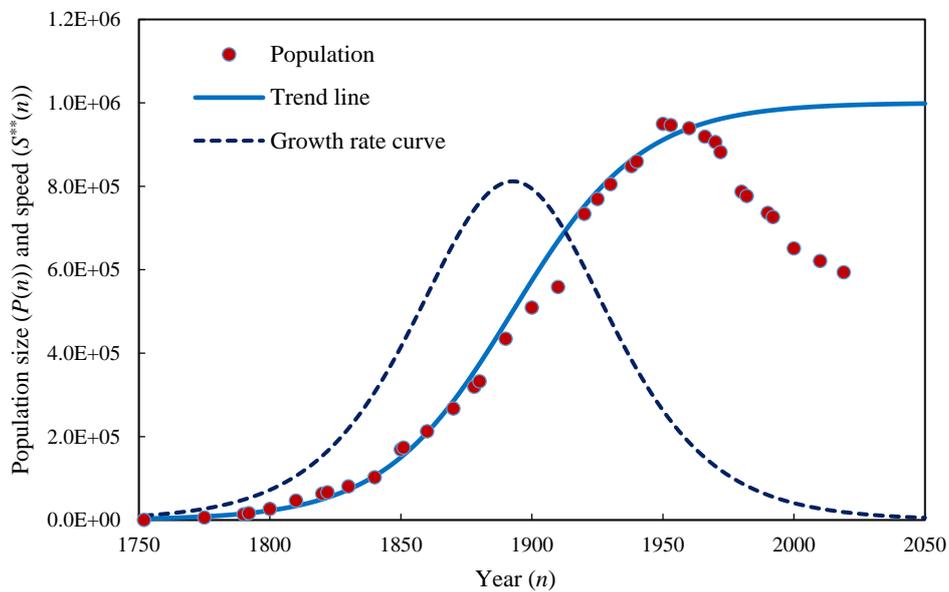

a. Population size and its growth speed

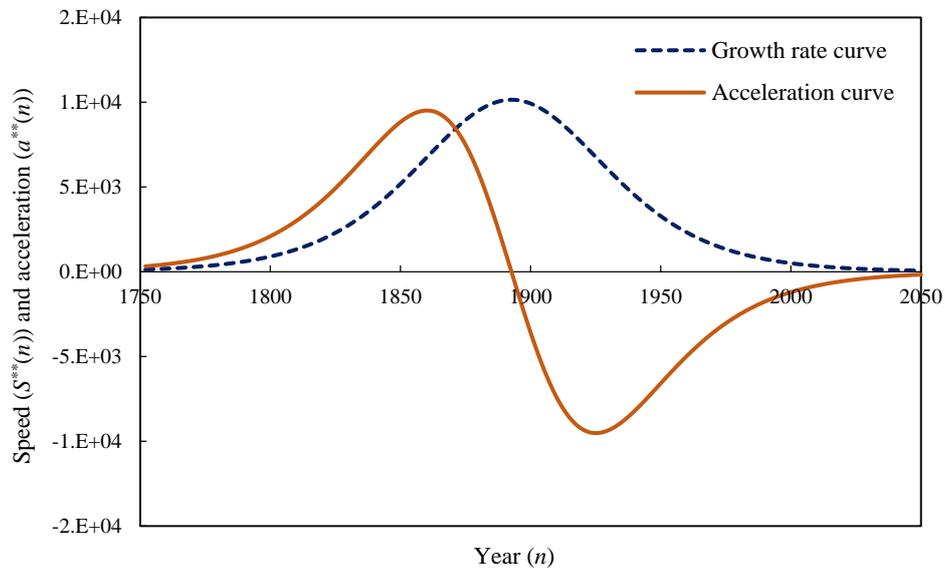

b. Population growth rate and acceleration

**Figure A2. Population size, population growth rate and acceleration of Baltimore, USA** (Note: Population data come from Shen (2002) and https://en.wikipedia.org/wiki/Baltimore. For facilitating intuitive comparison, the population growth rate is magnified 80 times in Figure A2 (a), and the population growth acceleration is magnified 60 times in Figure A2 (b).)